\documentclass[aps,prd,twocolumn,superscriptaddress,showkeys,showpacs]{revtex4}
\usepackage{mathptmx}
\usepackage[utf8]{inputenc}
\usepackage[T1]{fontenc}
\usepackage{slashed}
\usepackage{times}
\usepackage{amssymb,amsfonts,amsmath,amsthm}
\usepackage{dsfont,bbm}
\usepackage{dcolumn}
\usepackage{epsf}
\usepackage{graphicx}
\usepackage[caption=false]{subfig}
\usepackage{dsfont}
\usepackage{slashed}
\usepackage[active]{srcltx}
\usepackage[usenames]{color}
\begin{document}

\title{On the decomposition theorem for gluons}

\author{I.V.~Anikin}
\email{anikin@theor.jinr.ru}
\affiliation{Bogoliubov Laboratory of Theoretical Physics, JINR,
             141980 Dubna, Russia}
\author{A.S.~Zhevlakov}
\email{zhevlakov@theor.jinr.ru}
\affiliation{Bogoliubov Laboratory of Theoretical Physics, JINR,
             141980 Dubna, Russia}
\affiliation{Matrosov Institute for System Dynamics and
		Control Theory SB RAS, 664033 Irkutsk, Russia }

\begin{abstract}
Recently, the problem of spin and orbital angular momentum (AM) separation has widely been discussed.
Nowadays, all discussions about the possibility to separate the spin AM from the orbital AM in the gauge invariant manner
are based on the ansatz that the gluon field can be presented in form of
the decomposition where the physical gluon components are additive to the pure gauge gluon components,
{\it i.e.} $A_\mu = A_\mu^{\text{phys}}+A_\mu^{\text{pure}}$.
In the present paper,  we show that in the non-Abelian gauge theory 
this gluon decomposition has a strong mathematical evidence in the frame of the 
contour gauge conception. 
In other words, we reformulate the gluon decomposition ansatz as a theorem on decomposition and, then,
we use the contour gauge to prove this theorem.
In the first time, we also demonstrate that the contour gauge possesses the special kind of 
residual gauge related to the boundary field configurations and expressed in terms of the pure
gauge fields.
As a result, the trivial boundary conditions lead to the inference that the decomposition 
includes the physical gluon configurations only provided the contour gauge condition.
\end{abstract}
\pacs{13.40.-f,12.38.Bx,12.38.Lg}
\keywords{Contour Gauge, Path Group, Loop Group, Spin Angular Momentum, Orbital Angular Momentum.}
\date{\today}
\maketitle

%%%%%%%%%%%%%
\underline{\it Introduction.}
%%%%%%%%%%%%%%%%%%%%
One of interesting subjects of modern disputes in both the theoretical and experimental communities
is the possible separation of nucleon spin into the intrinsic spin and orbital angular momentum (AM)
of partons \cite{Ji:2020ena}.
Nowadays two concurrent  decompositions, as known as Jaffe-Manohar's decomposition (JM-decomposition)
\cite{Jaffe:1989jz} and Ji's decomposition (J-decomposition) \cite{Ji:1996ek}, have widely been discussed.
The JM-decomposition refers to a complete decomposition of the nucleon spin into
the spin and orbital parts of quarks and gluons individually.
While, the J-decomposition possesses the
gauge invariance by construction but, at the same time, it does not lead
to the separable quark and gluon contributions of spin and orbital AMs to the whole nucleon spin.

In \cite{Chen:2008ag,Chen:2009mr}, the gauge invariant analogue of JM-de\-com\-position
 has been proposed. Considering the Coulomb gauge condition, they have
 advocated that the gluon field can formally be presented as
\begin{eqnarray}
\label{Decom-1}
A_\mu(x) = A_\mu^{\text{phys}}(x)+A_\mu^{\text{pure}}(x).
\end{eqnarray}
It is worth to notice that this decomposition has being assumed as the fist-step ansatz in all existing discussions on
the gauge-invariant separation of spin AM from the orbital AM
(see, for example,  \cite{Ji:2009fu,Wakamatsu:2014zza,Wakamatsu:2014toa,Wakamatsu:2017isl,Lorce:2012ce,Lorce:2013gxa,Leader:2013jra,Wakamatsu:2011mb,Wakamatsu:2010cb,
Wakamatsu:2010qj,Zhang:2011rn,Bashinsky:1998if}).

In the Abelian $U(1)$ gauge theory the physical components $A_\mu^{\text{phys}}$ of Eqn.~(\ref{Decom-1})
correspond to the transverse components $A_\mu^{\perp}$ which are gauge invariant in contrast to the
longitudinal components $A_\mu^{L}$ associated with $A_\mu^{\text{pure}}$ which are gauge-transforming and
they should be eliminated by the gauge condition used in the Lagrangian approach.
As a result, in the Abelian theory the decomposition of Eqn.~(\ref{Decom-1}) is absolutely natural and
there is no doubt of its validity at all.

In the non-Abelian $SU(3)$ gauge theory, both the transverse and longitudinal components are gauge-transforming.
Hence, the mentioned decomposition is actually questioned regarding the definition of the physical components.
In particular, the use of covariant-type gauge conditions should inevitably lead to the inability to
separate the spin and orbital AMs in the gauge-invariant manner because the coordinate dependence of gluon configurations 
cannot be independently determined for every of components, see for example  \cite{Belitsky:2005qn}.   

Meanwhile, the decomposition plays a role of keystone in many discussions devoted to the
gauge-invariant separation of spin AM from orbital AM.

In the paper, we consider the decomposition of Eqn.~(\ref{Decom-1}) as a statement which must be proven, if it is possible,  
within the non-Abelian theory.
It turns out that the proof can be implemented and elucidated with the help of a contour gauge
which extends the standard local axial-type gauges and is free from the Gribov ambiquities \cite{ContourG1,ContourG2}.
Contrasting with the local gauge, in the contour gauge (which refers to the non-local type of gauges)
we should first fix the 
gauge orbit representative and then we search the local gauge condition which is suitable for a given gauge orbit representative.
The useful features of the contour gauge can be readily understood in the frame of the Hamiltonian formalism 
where the contour gauge condition defines the manifold surface crossing over a group orbit of the fiber uniquely
(see \cite{An-Sym} for further details).
an equation
As a new observation, we demonstrate that even in the contour gauge we can deal with a special kind of the residual ga\-uge freedom.
However, this residual gauge is located in the nontrivial boundary pure gauge configurations defined at infinity. 

Notice that the physical quantities do not depend on the choice of gauges, as it must be.  
The axial-type gauges are related to the certain fixed direction in a space.
In this case, the gauge independency should be treated as an independency with respect to the chosen direction
which is ensured by additional requirements \cite{Anikin:2009bf}. 
Moreover, the use of contour gauge implies that in the Hamiltonian formalism the gauge condition, as an additional condition,
can be completely resolved regarding the gauge function excluding the gauge transforms in the finite region of space.
Accordingly, in this meaning the physical observables considered in the contour gauge are gauge invariant ones by 
construction.  

%%%%%%%%%%%%%%
\underline{\it Local and non-local gauge transform conventions.}
%%%%%%%%%%%%%%%%%%%%%%%%%%%%%%%%%
Before going further, it is important to remind the convention system regarding the gauge transformations which match
the corresponding Wilson path functional (see \cite{Anikin:2016bor} for more details).
In what follows, for the sake of shortness, we say simply the Wilson line independently on the form of a path unless it
leads to misunderstanding.
For the non-Abelian gauge theory, let us now assume that the fermion and gauge fields are transformed as
\begin{eqnarray}
\label{Wl-GT-1}
&&\psi^{\theta}(x)=e^{+i\theta(x)} \psi(x)\equiv U(x) \psi(x) ,
\\
\label{Wl-GT-1-2}
&&A_{\mu}^\theta(x)= U(x)A_\mu(x) U^{-1}(x) + \frac{i}{g} U(x)\partial_\mu U^{-1}(x),
\end{eqnarray}
where $\theta=\theta^a T^a$ with $T^a$ being the generators of corresponding representations.
With the local transformations fixed as in Eqns.~(\ref{Wl-GT-1}) and (\ref{Wl-GT-1-2}),
we can readily see that the covariant derivative and
the gauge-invariant fermion string operator take the following forms:
\begin{eqnarray}
\label{Wl-GT-2}
%&&
i\, {\cal D}_{\mu} = i\, \partial_\mu + g A_\mu(x),
\quad
%\nonumber\\
%&&
\mathbb{O}^{\text{g.-inv.}}(x,y)=\bar\psi(y) [y \,;\, x]_{A} \psi(x)
\end{eqnarray}
with the Wilson line defined as
\begin{eqnarray}
\label{WL-def-1}
%&&
[x\, ; \, x_0 ]_A =
%\mathbb{P}\text{exp} \Big\{ ig\int_{x_0}^{x} d\omega_\mu A_\mu(\omega)\Big\}
%\nonumber\\
%&&\hspace{-0.3cm}
%\equiv 
\mathbb{P}\text{exp} \Big\{ ig\int_{P(x_0,x)} d\omega_\mu A_\mu(\omega)\Big\}
= {\bf g}(x | A)\equiv {\bf g}(P),
\end{eqnarray}
where $P(x_0,x)$ stands for a path connecting the starting $x_0$
and destination $x$ points in the Minkowski space.

Inserting the point $x_0$ in the Wilson line of the gauge-invariant string operator, see Eqn.~(\ref{Wl-GT-2}), we get that
\begin{eqnarray}
\label{WI-GT-3}
\mathbb{O}^{\text{g.-inv.}}(x,y)=\bar\psi(y) [y \,;\, x_0]_{A} [x_0 \,;\, x]_{A} \psi(x).
\end{eqnarray}
Eqn.~(\ref{WI-GT-3}) hints that the path-dependent non-local gauge transformation of fermions can be introduced in the form of
\begin{eqnarray}
\label{Wl-GT-4}
\psi^{{\bf g}}(x)  =  {\bf g}^{-1}(x | A) \psi(x),
\end{eqnarray}
where $\psi^{{\bf g}}(x)$ is nothing but the Mandelstam gauge-invariant 
fermion field $\Psi(x | A)$, modulo the global gauge transforms \cite{Mandelstam:1962mi,DeWitt:1962mg}.
This transformation leads, in turn, to (cf. Eqn.~(\ref{Wl-GT-1-2}))
\begin{eqnarray}
\label{A-trans-CG}
A_{\mu}^{{\bf g}}(x)= {\bf g}^{-1}(x | A)A_\mu(x) {\bf g}(x | A) + \frac{i}{g} {\bf g}^{-1}(x | A)\partial_\mu {\bf g}(x | A).
\end{eqnarray}
Therefore, we have the following correspondence between local and non-local gauge transformation
\begin{eqnarray}
\label{Corr-loc-nonloc}
 U(x) \Leftrightarrow {\bf g}^{-1}(x | A)
\end{eqnarray}
which is extremely important for the further discussions because the wrong correspondence results in the
substantially wrong conclusions (see for example \cite{Lorce:2012rr}).

%%%%%%%%%%%%%
\underline{\it Contour gauge conception.}
%%%%%%%%%%%%%%%%%%%%%%%%%%%%
In the paper, we make a reexamination of Eqn.~(\ref{Decom-1}) for clarification of the conditions
which provide the decomposition validity, unless the decomposition does not take place at all.
We intend to consider Eqn.~(\ref{Decom-1}) as a statement
which must be proven at least within the gauge condition that is more suitable for
a demonstration of Eqn.~(\ref{Decom-1}).
To this end, we adhere the contour gauge conception.

At the beginning, we remind that, within the Hamiltonian formulation of gauge theory \cite{Faddeev:1980be}, 
the extended functional integration measure over 
the generalized momenta, $E_i$, and coordinates, $A_i$, includes 
two kinds of the functional delta-functions. The first kind of delta-functions reflects the 
primary (secondary etc) constraints on $E_i$ and $A_i$,
while the second kind of delta-functions refers to the so-called additional constraints (or gauge conditions)
the exact forms of which have been dictated by the gauge freedom.
If the primary (secondary etc) constraints are needed to exclude the unphysical gauge field components,
the gauge conditions would allow, in the most ideal case, to fix the corresponding Lagrange factor related to the gauge orbit.       
Focusing on the Lagrangian formulation \cite{Faddeev:1973zb}, since the infinite volume of gauge orbit 
is factorized out in the functional measure over the gauge field components, the gauge conditions work for 
the elimination of unphysical gluon components.

In this connection, the contour gauge implies that in order to fix completely the gauge function (orbit representative) or
to eliminate the unphysical gluons, one can demand the Wilson path functional
between the starting point $x_0$ and the final destination point $x$, $P(x_0,x)$, to be equal to unity, {\it i.e.}
\begin{eqnarray}
\label{CG-1}
{\bf g}(x | A)=[x\, ; \, x_0 ]_A =\mathbb{I},
\end{eqnarray}
where the path $P(x_0,x)$ is now fixed and $x_0$ is a very special starting point that might depend on 
the destination point $x$, see also \cite{Weigert:1992my}.

In fact, the well-known axial gauge, like $A^+=0$, is a particular case of the most general non-local contour gauge
determined by the condition of Eqn.~(\ref{CG-1}) if the fixed path is the straightforward line
connecting $\pm\infty$ with $x$.

In the past, the contour gauge had been a subject of intense studies
(see, for example, \cite{ContourG1, ContourG2}).
The obvious preponderance of the contour gauge use
is that the quantum gauge theory is free from the Gribov ambiguities.
By construction, the contour gauge does not suffer from the residual gauge freedom 
and gives, from the technical point of view, the simplest way to fix totally the gauge in the finite space.
Briefly, within the contour gauge conception
we first fix an arbitrary point $(x_0, \textbf{g}(x_0))$
in the fiber ${\cal P}({\cal X}, \pi \,|\, G)$ \cite{Mensky:2012iy}. 
We define afterwards two directions: one direction is determined in the base ${\cal X}$
(where the direction is nothing but the tangent vector of a curve which
goes through the given point $x_0$), the other
direction -- in the fiber where the direction can be uniquely determined as the
tangent subspace related to the parallel transport.
These two directions form the horizontal vector (or direction) 
\begin{eqnarray}
\label{H-vector}
H_\mu = \partial_\mu -ig A^a_\mu(x) \, D^{a}
\end{eqnarray}
which is invariant under the structure group on the fiber by construction \cite{Mensky:2012iy}. 
In Eqn.~(\ref{H-vector}), $D^a$ denotes the corresponding shift generator 
along the group fiber and can be presented as ${\bf g} \partial/\partial {\bf g}$.
In ${\cal P}({\cal X}, \pi \,|\, G)$, the functional ${\bf g}(x | A)$ of Eqn.~(\ref{WL-def-1}) is a solution of 
the parallel transport equation given by
\begin{eqnarray}
\label{H-vector-2}
\frac{d x_{\mu}(s)}{d s} H_\mu(A) {\bf g}(x(s)| A) = 0
\end{eqnarray}
provided that $p(s)=\left( x(s), {\bf g}(x(s)) \right)$ is defined 
with the curve $x(s)\in {\cal X}$ parameterized by $s$. 
If we impose the condition of Eqn.~(\ref{CG-1}), to fulfil Eqn.~(\ref{H-vector-2}) 
we need to suppose either $A^a_\mu(x)=0$ (this is a trivial case) or 
$D^a \mathbb{I}=0$ (that is a natural requirement if $D^a={\bf g}\partial/ \partial{\bf g}$ as above).
  
We can thus uniquely define the point in the fiber bundle, ${\cal P}({\cal X}, \pi \,|\, G)$, which has the unique 
horizontal vector corresponding to the given tangent vector at $x\in {\cal X}$.
We remind that the tangent vector at the point $x$ is uniquely determined by the 
given path passing through $x$.   
That is, within the Hamilton formalism based on the geometry of gluons
the condition of Eqn.~(\ref{CG-1}) corresponds to the determining of the surface
on ${\cal P}({\cal X}, \pi \,|\, G)$. This surface is parallel to the base plane with the path and 
singles out the identity element, ${\bf g}=1$, in every fiber of ${\cal P}({\cal X}, \pi \,|\, G)$ \cite{An-Sym}. 

The contour gauge refers to the non-local class of gauges and
generalises naturally the familiar local axial-type of gauges. It is also worth
to notice that two different contour gauges can correspond to the same
local axial gauge where the residual gauge left unfixed \cite{Anikin:2010wz, Anikin:2015xka}.
This statement reflects the fact that, in contrast to the local axial gauge,
the contour gauge does not possess the residual gauge freedom in the finite region of a space.
However, as shown below, the boundary gluon configurations can generate the special class of the
residual gauges.

%%%%%%%%%%%
\underline{\it Contour gauge and the gluon field decomposition.}
%%%%%%%%%%%%%%%%%%%%%%%%%%%%%%
We are go\-ing over to the discussion of the contour gauge defined by the condition of Eqn.~(\ref{CG-1}).
Having used the path-dependent gauge transformations for gluons (see Eqn.~(\ref{A-trans-CG})), and
having calculated the derivation of the Wilson line \cite{Durand:1979sw}, we readily derive that
in the gauge $[x\, ; \, - \infty ]_A =\mathbb{I} $ the gluon field can be presented as the following decomposition
\begin{eqnarray}
\label{A-cg}
A^{\text{c.g.}}_\mu(x) = \int_{-\infty}^{ x} d\omega_\alpha G_{\alpha\mu}(\omega | A^{\text{c.g.}}) +
A^{\text{c.g.}}_\mu( x - n\, \infty ),
\end{eqnarray}
where $G_{\mu\nu}$ is the gluon strength tensor; 
the starting point is now equal to $-\infty$ and the path parametrization is given by
\begin{eqnarray}
\label{Path-par}
\omega \Big|_{x}^{ - \infty} = x - n \lim_{\epsilon\to 0} \frac{1-e^{-s\epsilon}}{\epsilon} \Big|_{s=0}^{s=\infty}.
\end{eqnarray}
This path parametrization includes the vector $n$ defined a given fixed direction.
As usual, the vector $n$ becomes a minus light-cone basis vector, $n=(0^+, n^-, {\bf 0}_\perp)$, within the approaches where
the light-cone quantization formalism has been applied.

Notice that the decomposition of Eqn.~(\ref{A-cg}) differs substantially from \cite{Hatta:2011zs}
by the absence of $\epsilon$-function. Indeed, the given contour gauge chooses either one $\theta$-function 
or the other, see \cite{Anikin:2015xka} for details.
 
From Eqn.~(\ref{A-cg}), we can see that the contour gauge allows the gluon field to be naturally separated on
the $G$-dependent and $G$-independent components. That is, instead of Eqn.~(\ref{A-cg})
it is instructive to write the separation as (cf. \cite{Bashinsky:1998if})
\begin{eqnarray}
\label{A-sep}
A^{\text{c.g.}}_\mu(x) = A_\mu(x | G) + A^{\text{b.c}}_\mu(-\infty),
\end{eqnarray}
where $A_\mu(x | G)$  is nothing but the
first term of Eqn.~(\ref{A-cg}) and
the boundary gluon configuration defined as 
$A^{\text{b.c}}_\mu(-\infty)\equiv A^{\text{c.g.}}_\mu(x - n\, \infty)$.
It is worth to notice that (a) the $G$-dependent configuration  $A_\mu(x | G)$ stems from the nontrivial deformation of a path
\cite{Durand:1979sw}; (b) the gluon separation presented by Eqn.~(\ref{A-sep}) 
resembles the equation of  \cite{Bashinsky:1998if} but differs slightly by meaning.

In the contour gauge, see Eqn.~(\ref{CG-1}), the boundary gluon configurations have to fulfil the condition as
\begin{eqnarray}
\label{bc-cond}
\mathbb{P}\text{exp} \Big\{ ig A^{\text{b.c}}_\mu(-\infty)\int_{-\infty}^x d\omega_\mu \Big\}=\mathbb{I}.
\end{eqnarray}
Therefore, since the integral over $d\omega_\mu$ in Eqn.~(\ref{bc-cond}) is divergent as $1/\epsilon$ at $\epsilon$ goes to zero,
the combination $n_\mu A^{\text{b.c}}_\mu(-\infty)$  should behave as $\epsilon^2$.
Indeed, the exponential function of Eqn.~(\ref{bc-cond}) reads (here, we deal with the space where the dimension is $D=4$)
\begin{eqnarray}
\label{bc-cond-2}
&&A^{\text{b.c}}_\mu(-\infty)\int_{-\infty}^x d\omega_\mu \equiv
A^{\text{b.c}}_\mu(x -\infty \,n)\int_{-\infty}^x d\omega_\mu=
\nonumber\\
%&&
%- \Big\{ \lim_{\epsilon\to 0} \Big(\frac{1}{\epsilon} \Big)^{-1}\Big\}
%A^{\text{b.c}}_\mu\big(-\frac{x}{\infty} + n\Big) \int_{-\infty}^x d\omega_\mu=
%\nonumber\\
&&\Big\{ \lim_{\epsilon\to 0} \Big(\frac{1}{\epsilon} \Big)^{-1}\Big\}
A^{\text{b.c}}_\mu(n) \, n_\mu  \Big\{ \lim_{\delta\to 0} \frac{1}{\delta}\Big\} = 0.
\end{eqnarray}
Hence, the boundary gluon configurations obey the transversity condition as
\begin{eqnarray}
\label{bc-cond-3}
n_\mu(\theta_i,\, \varphi) \, A^{\text{b.c}}_\mu\big (n(\theta_i,\, \varphi) \big)  = 0.
\end{eqnarray}
Here, since the vector $n$ defines the fixed direction it is more convenient to use the spherical co-ordinates in the Euclidean space
(or the pseudo-spherical system in the Minkowski space) where the vector $n$ depends on the angle co-ordinates
$(\theta_i, \varphi)$ only, see below.
If the space dimension is $D>4$, the transversity condition of Eqn.~(\ref{bc-cond-3}) is not necessary to fulfil the contour gauge
condition.

We are in position to show that in the contour gauge the boundary gluon configurations have the only form of the pure gauge
configurations. 
First of all, the starting point $x_0$ plays the special role in the
considered formalism because all the paths originate from this point and the base ${\cal X}$ touches the
principle fiber bundle ${\cal P}$ only at this point in the general path-dependent gauge by construction.

\begin{figure}[t]
\centerline{\includegraphics[width=0.25\textwidth]{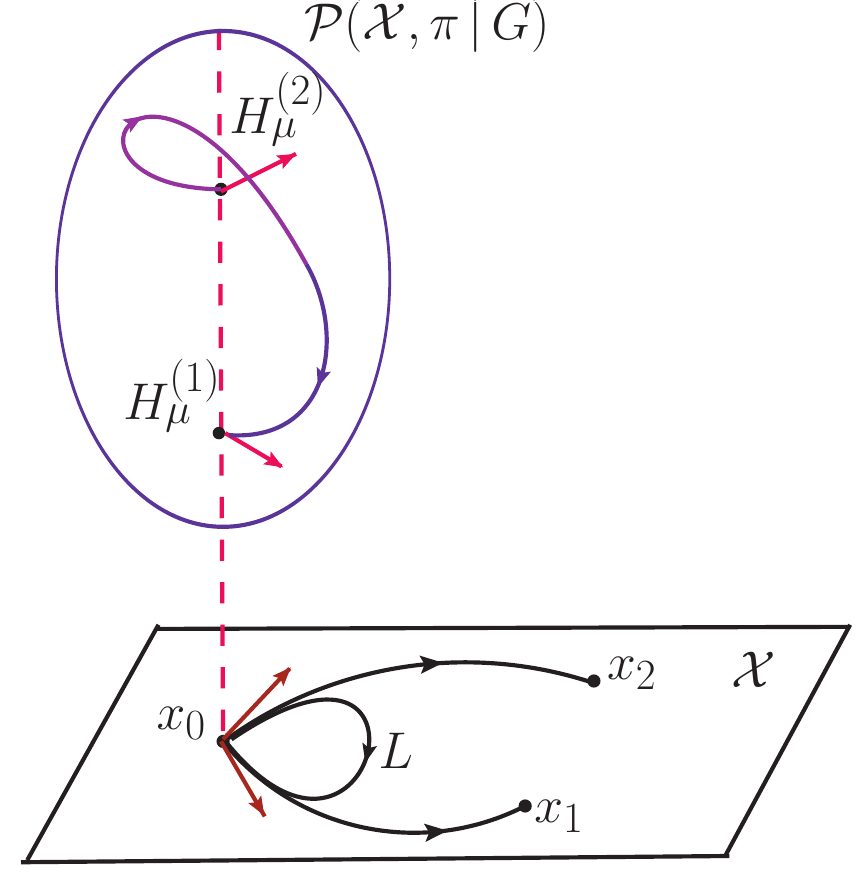}}
%\vspace{-0.5cm}
\caption{The holonomy: $H^{(i)}_\mu$ with $i=1,2$ denote the corresponding horizontal vectors defined on the 
given fiber of ${\cal P}$. } 
\label{Fig-1}
\end{figure}

Let us consider the point $x_0$ where, say, two different paths are started, see Fig.~(\ref{Fig-1}).
This starting point has two tangent vectors associated with $P(x_0, x_1)$ and $P(x_0, x_2)$.
In its turn, every of tangent vectors has the unique horizontal vector $H_\mu^{(i)}$ defined in the fiber.  
Then, making use of Eqn.~(\ref{CG-1}) we can obtain that
\begin{eqnarray}
\label{bc-cond-4_1}
&&\mathbb{P}\text{exp} \Big\{ ig \int_{L(x_0)} d\omega_\mu A_\mu(\omega)\Big\}=\mathbb{I}
\quad \text{and} \quad
\\
&&
\label{bc-cond-4_2}
\mathbb{P}\text{exp} \Big\{ ig \int_{\Omega} d\omega_\mu \wedge d\omega_\nu G_{\mu\nu}(\omega)\Big\}
=\mathbb{I},
\end{eqnarray}
where $L(x_0)$ implies the loop with a basepoint $x_0=-\infty$ and $\Omega$ is the corresponding surface related to the
loop $L(x_0)$, see Fig.~\ref{Fig-1}.
%
%
%%%%%%%%%%%%%%%%%%%%%%%%%%%%% FIGURE %%%%%%%%%%%%%%%%%%%%%%%%%%%%%%%%
%%%%%%%%%%%%%%%%%%%%%%%%%%%%%%%%%%%%%%%%%%%%%%%%%%%%%%%%%%%%%%%%%%%%%%%
%
Hence, we directly get that
\begin{eqnarray}
\label{bc-cond-5}
A_\mu(\omega) = \frac{i}{g}\,U(\omega) \partial_\mu U^{-1}(\omega)
\end{eqnarray}
from Eqn.~(\ref{bc-cond-4_1}), and  $G_{\mu\nu}(\omega)=0$ from  Eqn.~(\ref{bc-cond-4_2}) after the Stocks theorem has been used.

In the path group theory it states that any loop as a element of the loop subgroup can homotopically be transformed to the
"null element" which is, in our case, the basepoint $x_0=-\infty$.
As a result, the pure gauge representation of Eqn.~(\ref{bc-cond-5})
is valid for the boundary configurations  as well, {\it i.e.} we have
\begin{eqnarray}
\label{bc-cond-6}
A^{\text{b.c.}}_\mu(x_0) = \frac{i}{g}\,U(x_0) \partial_\mu U^{-1}(x_0).
\end{eqnarray}
Finally, combining Eqns.~(\ref{A-sep}) and (\ref{bc-cond-6}), we have proved that
in the contour gauge the gluon field can indeed be presented as the following decomposition
\begin{eqnarray}
\label{conc-1}
A^{\text{c.g.}}_\mu(x) = A_\mu (x | G)  + \frac{i}{g}\,U(x_0) \partial_\mu U^{-1}(x_0)\Big|_{x_0=-\infty},
\end{eqnarray}
where both terms are perpendicular to the chosen direction vector $n_\mu$.

\begin{figure}[t]
\centerline{\includegraphics[width=0.2\textwidth]{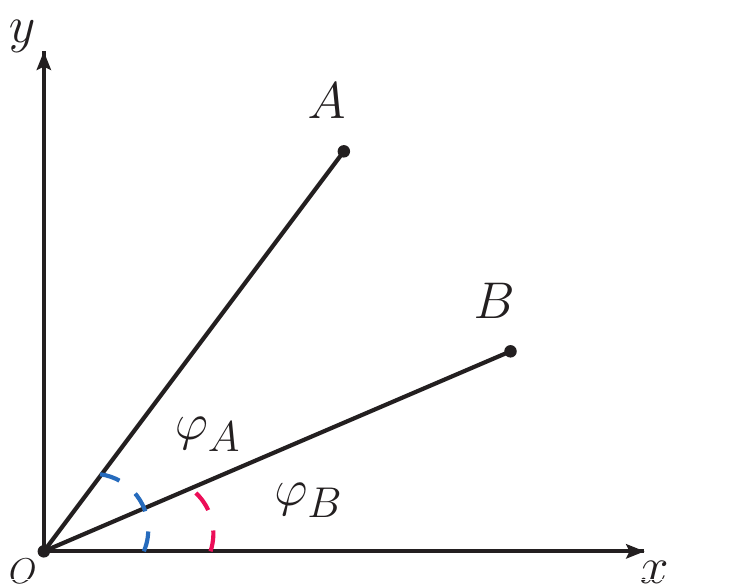}}
%\vspace{-0.5cm}
\caption{The angle independence of the starting point $O$: $A=(R_A, \varphi_A)$ and 
$B=(R_B, \varphi_B)$; $\lim_{R_{A}\to 0} A=\lim_{R_{B}\to 0} B=O$  where 
$O=(0\cdot\cos\varphi_i, 0\cdot\sin\varphi_i)$ with $i=A, B$.}
\label{Fig-2}
\end{figure}

Eqn.~(\ref{conc-1}) shows that the residual gauge of contour gauge is entirely located at the boundary.
Indeed, in order to understand the nature of the residual gauge associated with the boundary gluon configurations within
the contour gauge, we consider the simplest and illustrative example of $\mathbb{R}^2$ where $A$ and $B$ have the same starting point $O$,
see Fig.~\ref{Fig-2}. It is more convenient to work with the spherical system, {\it i.e.} 
$A(R_A, \varphi_A)\equiv (R_A \cos\varphi_A, R_A\sin\varphi_A)$ etc. 
If the radius vectors of both $A$ and $B$ differ from zero even infinitesimally, we can distinguish these two vectors.
However, if $R_A=R_B=0$, the starting point $O$ loses information on the vectors $A$ and $B$ because of
$O=(0\cdot\cos\varphi_A, 0\cdot \sin\varphi_A)=(0\cdot \cos\varphi_B, 0\cdot \sin\varphi_B)$. Notice that
in general case the angles can be arbitrary ones.
In this sense, we say that the starting point $O$ is the angle independent point.

Since $x_0=\lim_{R\to 0} X(R, \theta_1, \theta_2, \varphi)$, we have
\begin{eqnarray}
\label{res-g-1}
A^{\text{b.c.}}_\mu\big(  \bar\epsilon n(\theta_i, \varphi) \big) =\frac{i}{g}\,
U\big( \bar\epsilon n(\theta_i, \varphi) \big)
\partial_\mu U^{-1}\big( \bar\epsilon n(\theta_i, \varphi)  \big),
\end{eqnarray}
where $\bar\epsilon\to -\infty$ and
$\theta_i$ and $\varphi$ are not fixed ensuring the residual gauge freedom in the similar way as demonstrated in Fig.~\ref{Fig-2}.
%
%
%%%%%%%%%%%%%%%%%%%%%%%%%%%%% FIGURE %%%%%%%%%%%%%%%%%%%%%%%%%%%%%%%%
%%%%%%%%%%%%%%%%%%%%%%%%%%%%%%%%%%%%%%%%%%%%%%%%%%%%%%%%%%%%%%%%%%%%%%%
%

The most convenient way to fix the residual gauge freedom is to assume that
(see for instance \cite{Anikin:2010wz, Anikin:2015xka})
\begin{eqnarray}
\label{res-g-2}
A^{\text{b.c.}}_\mu\big( \bar\epsilon n(\theta_i, \varphi) \big)\equiv
A^{\text{pure}}_\mu\big( \bar\epsilon n(\theta_i, \varphi) \big)=0.
\end{eqnarray}
In this case, the decomposition presented by Eqn.~(\ref{Decom-1}) becomes a trivial one.

Thus, we have demonstrated that the contour gauge use gives the most natural decomposition
of gluon fields on the $G$-dependent gluon component, which can be called as the physical one,
and the unphysical gluon component related to the pure gauge configuration.
Moreover, in the finite domain of space the contour gauge does not suffer from the residual gauge,
while the remaining (possible) residual gauge has entirely been isolated on the infinite boundary
of a given space.

%%%%%%%%%
\underline{\it Local and non-local gauge matching.}
%%%%%%%%%%%%%%%%%%%%%%
Since the cont\-our ga\-uge as a non-local kind of gauges generalizes the standard local gauge of axial type,
it is worth to discuss shortly the correspondence between the local and non-local gauge transforms.
As mentioned, the local axial-type gauge suffers from the residual gauge transformations.
While the non-local contour gauge fixes all the gauge freedom in the finite space
provided the infinite starting point $x_0=-\infty$.
Indeed, if we consider the local axial gauge, $A^{+,\, \theta}(x)=0$, as an equation on the gauge function $\theta(x)$, see
Eqn.~(\ref{Wl-GT-1-2}), we can readily derive that the solution of this equation takes the form of
\begin{eqnarray}
\label{Sol-gauge}
&&{\cal U}_0(x^-, \tilde x)=C(\tilde x) \, \overline{\cal U}(x^-, \tilde x),
\nonumber\\
&&
\overline{\cal U}(x^-, \tilde x)= \mathbb{P}\text{exp}
\Big\{  - ig\int_{x_0^-}^{x^-} d\omega^- A^+(\omega^-, \tilde x)\Big\},
\end{eqnarray}
where $\tilde x = (x^+,0^-, {\bf x}_\perp)$, $x^-_0$ is fixed and $C(\tilde x)$ is an arbitrary function which does not depend on
$x^-$ and is given by
\begin{eqnarray}
\label{C-norm}
{\cal U}_0(x^- = x^-_0, \tilde x)=C(\tilde x).
\end{eqnarray}
The arbitrariness of $C$-function also reflects the fact we here deal with an arbitrary fixed starting point $x_0$.
We then come to the residual gauge freedom requiring both $A^{+,\, \theta}(x)=0$ and $A^{+}(x)=0$, we have
\begin{eqnarray}
\label{Sol-gauge-2}
&&{\cal U}_0(x^-, \tilde x)\Big|_{\overline{\cal U}=1}\equiv {\cal U}^{\text{res}}(\tilde x)=
C(\tilde x)\equiv
e^{i\,\tilde\theta(\tilde x)}.
\end{eqnarray}
One can see that the function $C$ determines the residual gauge transforms.

The non-local contour gauge extends the local axial-type gauge
and demands that the full integral in the exponential of Eqn.~(\ref{CG-1})
has to go to zero 
\footnote{In the local gauge, the corresponding exponential disappears thanks to
the nullified integrand $A^+=0$}. 
Within the Hamiltonian formalism, with the help of contour gauge,
the residual gauge function $\tilde\theta(\tilde x)$ can be related to the configurations $A^-$ and $A^i_\perp$
which are also disappeared eliminating the whole gauge freedom (see \cite{Anikin:2016bor,Belitsky:2002sm} for details).
That is, if we restore the full path in the Wilson line for a given process, we can derive that
\begin{eqnarray}
\label{C-full}
&&C(\tilde x)=\tilde C(x^+_0, x^-_0, {\bf x}^\perp_0)
%\nonumber\\
%&&\times
\mathbb{P}\text{exp}
\Big\{ ig\int_{{\bf x}_0^\perp}^{{\bf x}^\perp} d \omega_\perp^i A^i_\perp(x^+_0, x^-_0, \omega_\perp)\Big\}
\nonumber\\
&&\times  
\mathbb{P}\text{exp}
\Big\{ -ig\int_{x_0^+}^{ x^+} d \omega^+ A^-(\omega^+, x^-_0, {\bf x}_\perp)\Big\}
\end{eqnarray}
and, then, making use of the corresponding contour gauge we get that
\begin{eqnarray}
\label{C-full-2}
C(\tilde x)\Big|_{\text{c.g.}}=\tilde C(x^+_0, x^-_0, {\bf x}^\perp_0).
\end{eqnarray}
Eqn.~(\ref{C-full-2}) means that there is no the gauge freedom at all. 
The exact value of the fixed starting point $x_0$ depends on the process under 
our consideration \cite{Belitsky:2002sm,Burkardt:2012sd}. 

%%%%%%%%%
\underline{\it Non-zero boundary gluon configurations.}
%%%%%%%%%%%%%%%%%%%%%%
Let us study the influence of non-zero boundary gluon configurations on the definitions of different parton distributions.
We first emphasize that our decomposition of Eqns.~(\ref{A-cg}) and (\ref{A-sep}) relates in the meaning to the
decomposition of \cite{Bashinsky:1998if}. Indeed, we are able to rewrite Eqn.~(\ref{A-sep}) as
(here, we use the limit of $\bar\epsilon\to - \infty$)
\begin{eqnarray}
\label{A-dec-BJ}
%&&
\tilde A^{\text{l.c.}}_\mu(k^+; \tilde x) = G_\mu (k^+; \tilde x)  +
%\nonumber\\
%&&
\delta(k^+) A^{\text{b.c.}}_\mu\big( \bar\epsilon n^-(\pi/4, 0, 0) ; \tilde x \big), 
\end{eqnarray}
where
the light-cone gluon field $\tilde A^{\text{l.c.}}_\mu$ is the Fourier image of $A^{\text{l.c.}}_\mu$ with respect to $x^-$ only,
{\it i.e.}
\begin{eqnarray}
\label{F-im}
A^{\text{l.c.}}_\mu (x^-; \tilde x)\stackrel{{\cal F}}{=} \tilde A^{\text{l.c.}}_\mu(k^+; \tilde x),
\end{eqnarray}
and, therefore, we have  
\begin{eqnarray}
\label{M-BJ}
&&G^\mu (k^+; \tilde x) \stackrel{{\cal F}}{=} \int_{-\infty^-}^{ x^-} d\omega^- G^{ + \mu}(\omega^-, \tilde x | A^{\text{c.g.}})
\nonumber\\
&&\delta(k^+)\,A^{\text{b.c.}}_\mu \big( \bar\epsilon n^- ; \tilde x \big)\stackrel{{\cal F}}{=}
A^{\text{b.c.}}_\mu\big( \bar\epsilon n^- ; \tilde x \big).
\end{eqnarray}
Here, we underline that Eqns.~(\ref{A-dec-BJ}), as well as  Eqn.~(\ref{A-sep}), has been derived by direct solution of 
the contour gauge requirement, see Eqn.~(\ref{CG-1}).   
As mentioned, the important finding of the present paper is that despite the contour gauge fixes the whole gauge freedom in the finite domain
of space, it is still possible to deal with the residual gauge which is, however, located at the boundary field configurations only. 
The non-trivial topological effects due to the boundary field configurations are forthcoming in the further our studies.

In \cite{Bashinsky:1998if}, the representation that is similar to our Eqn.~(\ref{A-dec-BJ}) has rather been guessed in the local axial gauge,
$A^+=0$, where the corresponding residual gauge freedom is incorporated into the inhomogeneous term with $\delta(k^+)$.
In turn, the gauge $A^+=0$ with the fixed residual gauge
freedom in the finite domain of space is actually identical to the unique contour gauge \cite{Anikin:2015xka}.

In the frame of the path group formalism, we have the following path-dependent transformation,
which  generates the usual translation transformation,
\begin{eqnarray}
\label{PG-tran-1}
\big[ \mathbb{U}^{P(x, x+y)}\psi \big](x) = [x+y\, ; \, x ]^{-1}_A \, \psi(x+y),
\end{eqnarray}
where $\psi(x)$ belongs to the spinor fundamental representation and has defined on
the Minkowski space $M=P/L$ ($P$ denotes the corresponding path group,
$L$ stands for the loop subgroup of $P$) as a invariant function of the conjugacy classes,
{\it i.e.} $\psi(x)={\bf g}(p)\Psi(p)$ with $p=(x, {\bf g}) \in{\cal P}$ \cite{Mensky:2012iy}.
Besides, in Eqn.~(\ref{PG-tran-1}) the operator $\mathbb{U}$ which acts on the spinor manifold has
the form of
\begin{eqnarray}
\label{U-op-1}
\mathbb{U}^{P(x, x+y)} = \mathbb{P}\text{exp} \Big\{- ig\int_{x}^{x+y} d\omega_\mu {\cal D}_\mu \Big\}.
\end{eqnarray}

In the contour gauge where the Wilson line of Eqn.~(\ref{PG-tran-1}) is fixed to be equal to unity,  the 
transport operator $\mathbb{U}_q (y)$ takes the trivial form of
\begin{eqnarray}
\label{U-op-2}
\mathbb{U}^{P(x, x+y)} \Big|_{\text{c.g.}}= \mathbb{P}\text{exp} \Big\{- ig\int_{x}^{x+y} d\omega_\mu \partial_\mu \Big\}.
\end{eqnarray}
This operator does not include any information on the boundary configurations even if, say, $y\to\pm \infty$
because the boundary field configurations obey Eqn.~(\ref{CG-1}) too.
Moreover, in our case, the Wilson line of Eqn.~(\ref{PG-tran-1})
is set to unity due to the nullified integrand, $A^+=A^-=0$, and the nullified integral over $A_\perp$.

Hence, if we introduce the quark-gluon operators, forming the spin and orbital AM,
as the residual-gauge invariant operators, we have to use the covariant derivative as
$i\,{\cal D}^{\text{b.c.}}_\mu = i\,\partial_\mu  + g A^{\text{b.c.}}_\mu(-\infty)$.
In this sense, our results and the results of \cite{Bashinsky:1998if} are not much at variance .
For example, we readily obtain that 
\begin{eqnarray}
\label{BJ-1}
&&f_{L_q}(x) = {\cal N}\, \int_{-\infty}^{+\infty} dz^- e^{\,i x\,P^+ z^-} \int_{-\infty}^{+\infty} d^2{\bf y}_\perp 
\\
&&\times 
\langle P |
\bar \psi(y_\perp) \gamma^+ y^{[ i }_\perp \, i {\cal D}^{j ]}_{\text{b.c.}} \psi(y_\perp + z^-)
|P \rangle,
\nonumber 
\end{eqnarray}
where the antisymmetric combination $[i\, j]$ has been introduced  with $i, j= 1, 2$ and ${\cal N}$ is the normalization factor defined as in  \cite{Bashinsky:1998if}.

As above-mentioned, $f_{L_q}(x) $ as the physical quantity does not depend on the gauge choice.  
At the same time, the axial-type (local or non-local) gauges are correlated with the fixed direction
which is also necessary for the factorization procedure \cite{Anikin:2009bf}.
Therefore, we are able to treat the gauge independency in the meaning of an independency on the chosen direction
as well. 
In the frame of the Hamiltonian formalism, we assume that the gauge condition (or an additional condition)
can be completely resolved regarding the gauge function excluding the gauge transforms in the finite region.
In a sense, the physical quark-gluon operators, considered in the contour gauge, are ``gauge invariant'' by 
construction because we do not deal with any gauge transforms in the finite region due to the fixed gauge function $\theta_{\text{fix}}$
(as above discussed, due to ${\bf g}=1$ in the fiber for the whole base ${\cal X}$), see \cite{An-Sym} for details.  

%%%%%%%%%
\underline{\it Conclusions.}
%%%%%%%%%%%%%%%%%%%%%%
To conclude, we have expounded the useful correspondence between local and non-local gauges
which is extremely important to avoid the substantially wrong conclusions appeared in the literature. 

We have proposed the proof of the following statement which is valid in the non-Abelian theory:
in the contour gauge the gluon field can be presented in the form
of decomposition
on the gluon configuration $A_\mu(x | G)$ being the physical degree of freedom and
the pure gauge gluon configuration $A_\mu^{\text{pure}}(x_0)$ that is totally isolated on the boundary
and includes the special type of residual gauge freedom. 
We have demonstrated that the contour gauge condition cannot finally eliminate this, new-found, special residual gauge
the nature of which has been illustrated in detail.
  
In the case of the trivial boundary conditions, {\it i.e.}
$A^{\text{b.c.}}_\mu=0$, in the contour gauge 
the decomposition of Eqn.~(\ref{Decom-1})  does not make a sense in the non-Abelian theory 
because only the boundary gluon configurations can be presented as the pure gauge gluon configurations.  
Moreover, if the boundary configurations have been nullified, there is no the gauge freedom at all 
and, therefore, we deal with the gauge invariant operators by construction modulo the global gauge transformations that are 
not essential for the bilinear forms. 

As a last point, we want to mention that the gluon decomposition of \cite{Chen:2008ag}, which is formally similar to Eqn.~(\ref{Decom-1}),
has a status of the ansatz rather then a strong inference formulated and proven in our studies.
Moreover, it has a distinguished feature that
the gluon fields are separated into the physical and pure gauge gluon configurations 
before the gauge condition has been fixed. Hence, in this case, in order to formulate  
the ansatz they should demand to impose 
an addition requirement to extract $A^{\text{pure}}_\mu(x)$ which is finally defined by $G^{\text{pure}}_{\mu\nu}(x) = 0$.  
In its turn, this requirement appears naturally working within contour gauge conception, see Eqn.~(\ref{conc-1}). 
Eqn.~(\ref{BJ-1}), is formally not at odds with \cite{Chen:2008ag,Bashinsky:1998if} but,
in a sense,  we are in contradiction with \cite{Ji:1996ek,Wakamatsu:2010cb}.
%%%%%%%%%%%%%%
% Acknowledgements
%%%%%%%%%%%%%%
 
We thank C.~Lorce, 
D.G.~Pak, 
M.V.~Polyakov, 
L.~Szymanowski for useful discussions. IVA is grateful to O.V.~Teryaev
for fruitful comments on the early stage of the work.

\end{document}